\documentclass{article}
\usepackage[utf8]{inputenc}
\usepackage{csquotes}
\usepackage[english]{babel}
\usepackage[letterpaper,top=2cm,bottom=2cm,left=3cm,right=3cm,marginparwidth=1.75cm]{geometry}
\usepackage{amsmath}
\usepackage{graphicx}
\usepackage[colorlinks=true, allcolors=blue]{hyperref}
\usepackage{authblk}

\usepackage[normalem]{ulem} 
\newcommand{\stkout}[1]{\ifmmode\text{\sout{\ensuremath{#1}}}\else\sout{#1}\fi}

\newcommand{\mdsicu}[0]{\textit{MDS-ICU}}

\usepackage{graphicx}
\usepackage{siunitx}
\usepackage{xcolor}
\usepackage{comment}
\newcommand{\heading}[1]{\noindent\textbf{#1}}
\usepackage{graphicx}
\usepackage{amsmath}
\usepackage{pifont}
\usepackage{amsmath, amssymb, amsfonts, amsthm}
\usepackage{graphicx}
\usepackage{amsfonts}       
\usepackage{nicefrac}       
\usepackage{microtype}      
\usepackage{mathtools}
\usepackage{comment}
\usepackage{pbox}
\usepackage{textcomp}

\usepackage{multirow}
\usepackage{nameref}
\usepackage{textalpha}
\usepackage{booktabs} 
\usepackage{comment}
\usepackage{subfig}
\usepackage{threeparttable}
\usepackage{caption}
\usepackage{ragged2e} 

\usepackage{longtable}

\usepackage[
  backend=biber,
  style=nature,
  doi=false,
  url=false
]{biblatex}

\usepackage{tcolorbox}

\title{A Multimodal Deep Learning Framework for Predicting ICU Deterioration: Integrating ECG Waveforms with Clinical Data and Clinician Benchmarking}

\author[1]{Juan Miguel López Alcaraz}
\author[2]{Xicoténcatl López Moran}
\author[3]{Erick Dávila Zaragoza}
\author[4]{Claas Händel}
\author[5]{Richard Koebe}
\author[6]{Wilhelm Haverkamp}
\author[1,*]{Nils Strodthoff}

\affil[1]{AI4Health Division, Carl von Ossietzky Universität Oldenburg, Oldenburg, Germany}
\affil[2]{CAPASITS, Secretariat of Health of the State of Colima, Tecomán, México}
\affil[3]{Department of Interventional Cardiology, Hospital Puerta de Hierro Colima, Colima, México}
\affil[4]{Institute for Medical Informatics and Statistics, Kiel University and University Hospital Schleswig-Holstein, Kiel, Germany}
\affil[5]{Department of Anesthesiology, Intensive Care Medicine, Emergency Medicine, and Pain Therapy, Carl von Ossietzky Universität \& Klinikum Oldenburg, Oldenburg, Germany}
\affil[6]{Department of Cardiology, Angiology and Intensive Care Medicine, Charité Campus Mitte, German Heart Center of the Charité-University Medicine Berlin, Berlin, Germany}

\usepackage{booktabs}
\usepackage{comment}

\date{}

\begin{document}
\maketitle


\begin{abstract}
Artificial intelligence (AI) holds substantial promise for supporting clinical decision-making in intensive care units (ICUs), where timely and accurate risk assessment is essential. However, many existing predictive models remain narrowly scoped, focusing on isolated outcomes or limited data modalities, whereas clinicians routinely synthesize longitudinal clinical history, real-time physiological signals, and heterogeneous diagnostic information when assessing patient risk. To address this gap, we developed MDS-ICU, a unified multimodal machine learning framework that fuses routinely collected patient data—including demographics, biometrics, vital signs, laboratory values, ECG waveforms, surgical procedures, and medical device usage—to provide comprehensive predictive support throughout ICU stays. Using data from 63,001 samples across 27,062 patients in MIMIC-IV, we trained a deep learning architecture combining structured state-space (S4) encoders for ECG waveforms with multilayer perceptron (RealMLP) encoders for tabular features to jointly predict 33 clinically relevant outcomes spanning mortality, organ dysfunction, medication requirements, and acute deterioration events. The framework demonstrated strong discriminative performance, achieving AUROCs of 0.90 for 24-hour mortality, 0.92 for sedative administration, 0.97 for invasive mechanical ventilation, and 0.93 for coagulation dysfunction. Calibration analyses showed close alignment between predicted probabilities and observed outcomes, with ECG waveform inclusion yielding consistent improvements in both discrimination and predictive reliability. Direct comparison with clinicians and LLMs showed that model predictions alone outperformed clinicians' judgment and LLM predictions. Furthermore, providing model outputs as augmented decision support further increased predictive accuracy of both clinicians and LLMs. These results demonstrate that multimodal AI frameworks can deliver continuous, clinically meaningful risk stratification across diverse ICU outcomes while empirically showing that AI augments—rather than replaces—clinical expertise. This work establishes a scalable foundation for the integration of high-frequency physiological data into precision-oriented critical care decision support.

\end{abstract}

\section*{Introduction}

\heading{Prospects of AI in the ICU} Artificial intelligence (AI) is rapidly reshaping critical care medicine, particularly in the intensive care unit (ICU), where timely and accurate decisions are essential \cite{van2021moving}. Clinicians must continuously monitor complex physiological signals and respond to signs of deterioration, often under severe time pressure \cite{blythe2024clinician}. AI-driven systems have shown promise in enabling early detection of adverse events, improving triage, forecasting deterioration, and supporting diagnostic reasoning \cite{alcaraz2024mds}. By leveraging large-scale clinical records and real-time physiological data, these models can help mitigate the cognitive overload inherent to ICU environments \cite{8482421}. Despite this potential, current clinical AI systems frequently remain limited in scope.

\heading{Narrow scope of existing prediction models} Most existing ICU models focus on a narrow set of predefined outcomes, rely on short prediction horizons that fail to reflect long-term patient trajectories \cite{allam2021analyzing}, and underutilize rich multimodal clinical context \cite{acosta2022multimodal}. In particular, continuous physiological waveforms such as electrocardiograms (ECGs) are often excluded \cite{wornow2024ehrshot,yeche2021hirid}, despite evidence that they can outperform traditional clinical variables across prognostic and non-cardiovascular tasks \cite{alcaraz2024mds}. These signals provide high-resolution, real-time insight into patient physiology and are routinely collected as part of standard ICU monitoring \cite{goldstein1997intensive}, making their omission a notable limitation.

\heading{Lack of benchmarking datasets} Progress is further constrained by the lack of high-quality, open-access, multimodal ICU datasets. Widely used benchmarks \cite{wornow2024ehrshot,sheikhalishahi2020benchmarking,yeche2021hirid} often lack waveform data, temporal structure, or a diverse set of clinically meaningful prediction targets. Although recent efforts aggregate multiple data sources across large databases \cite{water2024yet,gupta2022extensive}, they still fall short in modality integration, task diversity, and representation of longitudinal patient trajectories. Crucially, most studies evaluate models exclusively against retrospective labels, without systematic comparison to clinician judgment or assessment of clinical significance \cite{binuya2022methodological}. As a result, strong predictive performance does not necessarily translate into actionable clinical value, limiting trust and deployment relevance.

\heading{Contributions} In this work, we introduce a generalist multimodal decision-support system for ICU care that integrates raw ECG waveforms with structured clinical data and explicitly benchmarks AI predictions against clinical judgment. Our contributions are threefold: (1) The construction and public release of a rich multimodal ICU dataset combining demographics, biometrics, vital sign and laboratory trends, surgical procedures, medical device usage, and raw ECG waveforms, together with 33 clinically meaningful deterioration targets spanning mortality, medication administration, organ dysfunction (Sepsis-3 SOFA), and acute clinical events. (2) A unified multimodal fusion framework that integrates high-frequency ECG waveforms with discrete clinical markers into a shared latent representation. This is achieved by combining statistical trend-based feature extraction for non-waveform time series, structured state-space models for ECG waveforms, and an enhanced deep tabular encoder. The architecture supports joint prediction across heterogeneous modalities, yielding high predictive performance (macro AUROC = 0.865) and strong calibration (Brier score = 0.0860). By comparison, the unimodal non-waveform model achieves 0.8583 AUROC and 0.0873 calibration, highlighting the added value of incorporating waveform information. (3) Clinician-centered benchmarks evaluating (i) clinician and LLM judgment versus model predictions where the proposed model outperforms both, and (ii) clinician/LLM judgment augmented with model outputs where on average, AI predictions increased the clinicians' predictive by a Youden index increase of 12\% and LLM performance by 16\%, demonstrating the model’s clinically meaningful benefit both alone and as augmented support. We envision this resource as both a clinically relevant benchmark and a foundation for reproducible research in ICU decision support. We refer to this dataset and framework as \textit{MDS-ICU} (Multimodal Decision Support in the ICU).

\begin{figure*}[!ht]
    \centering
    \includegraphics[width=\textwidth]{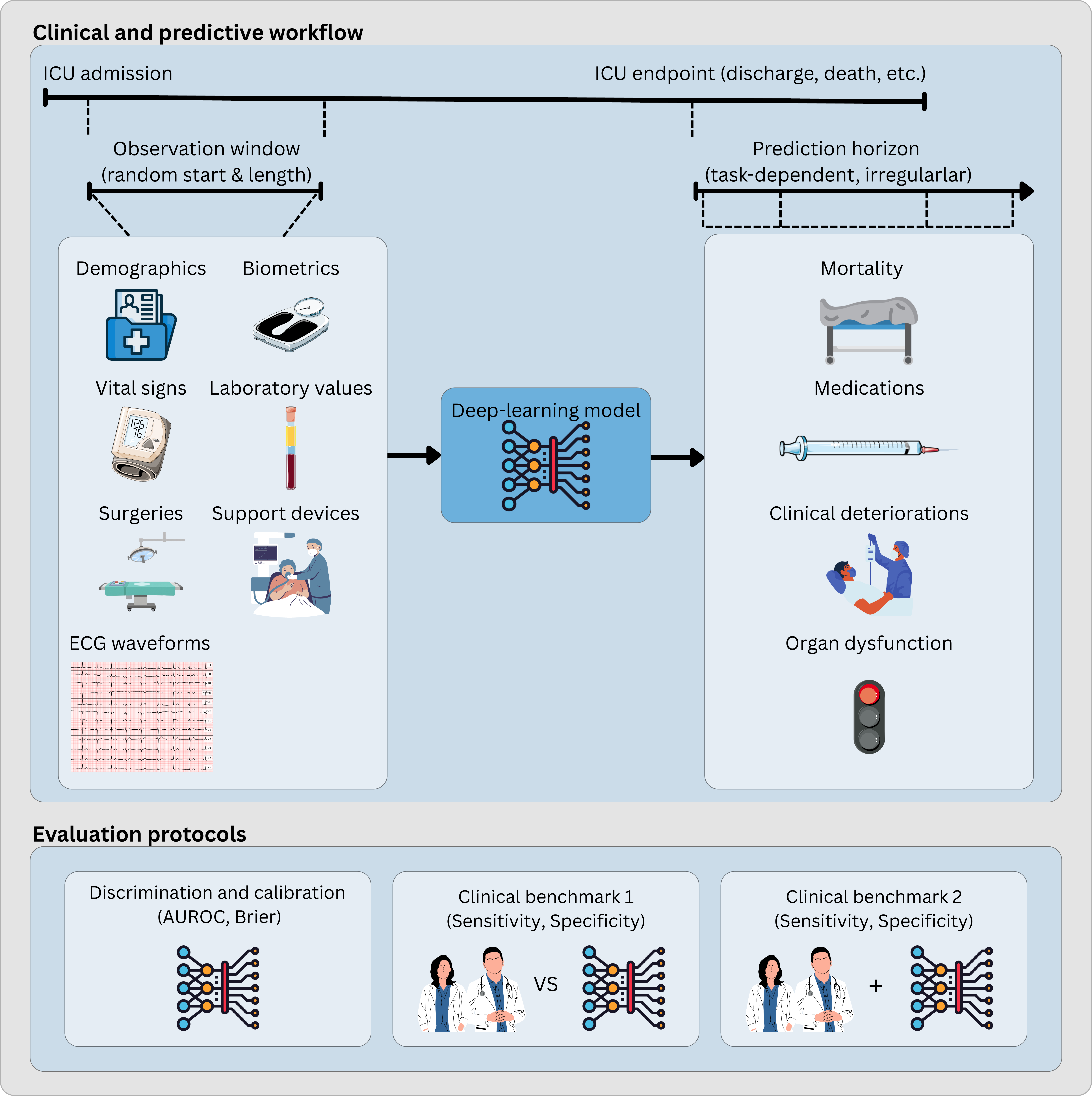} 
    \caption{Overview of the \mdsicu clinical and predictive modeling workflow. For each 10-second, 12-lead ECG recorded during an ICU stay, a corresponding observation window is constructed, spanning from ICU admission up to the ECG acquisition time. Within this observation window, multimodal clinical data, including demographics, vital signs, laboratory measurements, and biometrics, are extracted to form the feature set. These inputs feed into a unified model trained to perform a wide range of prediction tasks, such as clinical deterioration, early warning scores, medication administrations, discharge diagnoses, discharges, and mortalities. The dataset integrates and harmonizes structured clinical records and waveform data from MIMIC-IV, MIMIC-IV-ECG, and MIMIC-IV-ECG-ICD, enabling a comprehensive and temporally aligned view of each ICU episode. Rigorous preprocessing ensures the clinical plausibility and quality of extracted features, establishing a robust foundation for training generalizable and clinically useful decision support systems.}
    \label{fig:data_extraction_workflow}
\end{figure*}

\section*{Results}
\heading{Approach} We developed \mdsicu{}, a multimodal predictive framework for intensive care decision support that integrates 10-second 12-lead ECG waveforms with comprehensive tabular clinical data from the MIMIC-IV database. The dataset comprises 63,001 samples from 27,062 patients across 33,590 ICU stays, with each sample containing raw ECG signals and 801 temporal features derived from demographics, vital signs, laboratory measurements, and procedural context. We trained models to predict 33 clinically relevant labels spanning four categories: mortality at multiple time horizons (ICU, hospital, and 1-365 days), organ dysfunction (Sepsis-3 criteria), medication administration within 24 hours (13 drug classes), and acute clinical deterioration events (hypoxemia, mechanical ventilation, cardiac arrest, extracorporeal membrane oxygenation(ECMO)). Our approach employs S4 encoders for ECG waveforms and RealMLP for tabular features with late fusion, evaluated using stratified patient-wise splits and macro-averaged AUROC as the primary metric. We further benchmarked model predictions against assessments by ICU clinicians on a subset of cases, both independently and with AI-augmented decision support. Figure \ref{fig:data_extraction_workflow} illustrates the \mdsicu{} predictive workflow.

\begin{table}[ht!]
\centering
\caption{Descriptive statistics of the \mdsicu{} dataset at the sample level. Continuous variables are summarized as median (interquartile range) with units, and categorical or binary variables are reported as counts and percentages. Feature groups include demographics, biometrics, vital signs (cardiovascular, respiratory/ventilation, neurological, temperature), laboratory values (hematology, electrolytes/metabolic, renal/liver, other), surgeries, and mechanical ventilation. The ECG features are waveform-derived features and are only for reference of the waveforms in-hand, they were not used as input features in the model, all remaining features serve as input features for the model along with a 10s ECG waveform.}
\label{tab:descriptive_stats}
\small
\begin{tabular}{p{3.5cm} p{11cm}}
\hline
\textbf{Feature Group} & \textbf{Statistics} \\
\hline
Demographics &
Age (quantiles): 18--57: 26.34\%, 57--68: 25.39\%, 68--78: 24.30\%, 78--100: 23.98\% \par
Gender: Male 60.22\%, Female 39.77\% \par
Ethnicity: White 67.39\%, Other 16.51\%, Black 10.02\%, Hispanic 3.43\%, Asian 2.65\%
\\
\hline
Biometrics &
Height: 66 in (63--69), Weight: 169 lb (143--201), Body Mass Index: 27 kg/m$^{2}$ (23--31)
\\
\hline
Vital Signs &
\textbf{Cardiovascular:} Systolic Blood Pressure 114 mmHg (101--130), Diastolic Blood Pressure 58 mmHg (51--67), Mean Arterial Pressure 8.5 kPa (7--11) , Central Venous Pressure 11 mmHg (8--15), Heart Rate 84 bpm (72--99), Non-invasive Systolic Blood Pressure 115 mmHg (101--132), Non-invasive Diastolic Blood Pressure 63 mmHg (53--74) \par
\textbf{Respiratory and Ventilation:} Respiratory Rate 19 breaths/min (16--23), Tidal Volume 473 mL (400--545), Minute Volume 8.4 L/min (7.1--10.4), Positive End-Expiratory Pressure 5 cmH$_2$O (5--5), Peak Inspiratory Pressure 19 cmH$_2$O (13--24), Inspired Oxygen Fraction 50\% (40--60), Oxygen Flow 3 L/min (2--10), Oxygen Saturation 98\% (95--100), Apnea Interval 20 s (20--20) \par
\textbf{Neurological and Consciousness:} Glasgow Coma Scale Eye Spontaneously 0 (Spontaneously 0--To Speech 1), Glasgow Coma Scale Motor Obeys Commands 0 (Obeys Commands 0--Localizes Pain 1), Glasgow Coma Scale Verbal Confused 1 (Oriented 0--No Response-ETT 5), Level of Consciousness Alert 0 (Alert 0--Arouse to Stimulation 3), Goal Richmond Agitation-Sedation Scale Alert and calm 0 (Alert and calm 0--Awakens to voice 1), Richmond Agitation-Sedation Scale Alert and calm 4 (Alert and calm 4--Light sedation 6)  \par
\textbf{Temperature:} Temperature 98.3 °F (97.7--98.9)
\\
\hline
Laboratory Values &
\textbf{Hematology:} White Blood Cells 10.9 x10$^3$/µL (8--15), Red Blood Cells 3.36 x10$^6$/µL (2.91--3.86), Hemoglobin 10.0 g/dL (8.7--11.5), Hematocrit 30.4\% (26.7--34.7), Platelet Count 183 x10$^3$/µL (129--252), Basophils 0.02 x10$^3$/µL (0--0.04), Eosinophils 0.6 x10$^3$/µL (0.1--1.6), Lymphocytes 10.0\% (5.6--16.4), Neutrophils 81.0\% (73.8--87) \par
\textbf{Electrolytes and Metabolic:} Sodium 139 mmol/L (136--142), Potassium 4.1 mmol/L (3.8--4.5), Chloride 104 mmol/L (99--108), Calcium 8.4 mg/dL (7.9--8.9), Ionized Calcium 1.13 mmol/L (1.08--1.18), Magnesium 2.1 mg/dL (1.9--2.3), Phosphate 3.5 mg/dL (2.8--4.3), Bicarbonate 24 mmol/L (21--27) \par
\textbf{Renal and Liver:} Creatinine 1.1 mg/dL (0.8--1.8), Blood Urea Nitrogen 23 mg/dL (15--40), Bilirubin Total 0.7 mg/dL (0.4--1.3), Bilirubin Direct 1.7 mg/dL (0.6--4.5), Albumin 3.0 g/dL (2.5--3.4), Alkaline Phosphatase 87 U/L (63--132), Alanine Aminotransferase 30 U/L (17--67) \par
\textbf{Other:} C-Reactive Protein 93.7 mg/L (36.7--179.9), Troponin T 0.1 ng/mL (0.01--0.57), Carbon Dioxide Production 188 mL/min (154--232), End-Tidal Carbon Dioxide 35 mmHg (30--40), pH 7.37 (7.3--7.43), Oxygen Saturation 94\% (74--97)
\\
\hline
Surgeries &
Cardiac 12,100 (19.21\%), General 5,260 (8.35\%), Neurosurgery 3,547 (5.63\%), Vascular 1,467 (2.33\%), Thoracic 1,079 (1.71\%), Plastic 116 (0.18\%)
\\
\hline
Mechanical Ventilation &
Invasive 33,989 (53.95\%), Non-invasive 2,758 (4.38\%)
\\
\hline
ECG Features &
QRS Duration 96 ms (86--113), P-wave Duration 104 ms (72--120), RR Interval 740 ms (612--882), JT Interval 288 ms (254--322), P Axis 60° (40--86), QRS Axis 13° (–18--48), T Axis 48° (11--84)
\\
\hline
\end{tabular}
\end{table}

\heading{Descriptive summary} Table \ref{tab:descriptive_stats} provides descriptive statistics of the \mdsicu{} dataset across major clinical feature groups, including demographics, biometrics, vital signs, laboratory values, surgical interventions, mechanical ventilation, and ECG-derived features. The cohort shows a broadly uniform age distribution across quartiles (18–57, 57–68, 68–78, and 78–100 years, each contributing ~24–26\%), with a predominance of male patients (60.22\%). Ethnically, the population is mainly white (67.39\%), followed by other (16.51\%), black (10.02\%), hispanic (3.43\%), and asian (2.65\%) groups. Biometrics indicate a median body mass index in the overweight range (27 kg/m$^2$), while vital signs reflect typical ICU-level cardiovascular, respiratory/ventilatory, neurological, and temperature profiles. Laboratory measurements span hematological, metabolic, renal, hepatic, and inflammatory markers, capturing wide interquartile ranges consistent with acute critical illness. Procedural data reveal that cardiac and general surgeries were the most frequent, over half of the samples involved invasive mechanical ventilation, and ECG waveform-derived features exhibit physiologically plausible distributions. Collectively, these statistics highlight the clinical heterogeneity and demographic diversity of the ICU population represented in the \mdsicu{} dataset.

\subsection*{Predictive performance}

\begin{table*}[ht!]
\centering
\footnotesize
\caption{Comparative predictive performance (AUROC). Predictive performance of the multimodal fusion architecture integrating ECG waveforms and routine clinical data compared with a clinical-data-only baseline across 33 ICU outcomes spanning mortality, medication administration, clinical deterioration, and organ dysfunction. Outcome counts and prevalence are reported in parentheses. The best-performing model is highlighted in bold and underlined; when the alternative model is not statistically significantly worse, it is also shown in bold. Rows where only the multimodal model is highlighted indicate prediction tasks where the integration of ECG waveforms provides a significant performance advantage. Exemplary tasks considered for the clinical benchmark are marked with an asterisk.}
\label{tab:performance}
\begin{tabular}{lcc}
\hline
\textbf{Prediction target} &
\textbf{Multimodal} &
\textbf{Unimodal} \\
\textbf{(counts, prevalence)} &
\textbf{(ECG+routine data)} &
\textbf{(routine data)}\\
& \textbf{S4+RealMLP} & \textbf{RealMLP}\\
\hline
\multicolumn{3}{l}{\textbf{\textit{Mortality (different horizons)}}} \\
ICU mortality (7,028; 11.16\%) & \textbf{\underline{0.8809}} & 0.8459\\
Stay mortality (9,443; 14.99\%)${}^\ast$ & \textbf{\underline{0.8561}} & 0.8356\\
1-day mortality (1,822; 2.89\%) & \textbf{\underline{0.9009}} & \textbf{0.8932}\\
2-day mortality (2,744; 4.36\%) & \textbf{\underline{0.8834}} & \textbf{0.8809}\\
7-day mortality (5,968; 9.47\%) & \textbf{\underline{0.8645}} & 0.8416\\
28-day mortality (11,128; 17.66\%) & \textbf{\underline{0.8609}} & 0.8349\\
90-day mortality (15,097; 23.96\%) & \textbf{\underline{0.8495}} & 0.8237\\
180-day mortality (17,681; 28.06\%) & \textbf{\underline{0.8393}} & 0.8166\\
1-year mortality (20,752; 32.94\%) & \textbf{\underline{0.8320}} & 0.8109\\
\hline
\multicolumn{3}{l}{\textbf{\textit{Medications (administration within the next 24 hours)}}} \\
Crystalloids (57,752; 91.67\%) & \textbf{0.8811} & \textbf{\underline{0.8883}}\\
Electrolytes (33,568; 53.28\%) & \textbf{0.8017} & \textbf{\underline{0.8101}}\\
Antibiotics (37,136; 58.95\%) & \textbf{\underline{0.8570}} & \textbf{0.8562}\\
Vasopressors (21,012; 33.35\%)${}^\ast$ & \textbf{\underline{0.8854}} & \textbf{0.8781}\\
Inotropes (5,746; 9.12\%) & \textbf{0.8732} & \textbf{\underline{0.8738}}\\
Antiarrhythmics (6,663; 10.58\%) & \textbf{\underline{0.7909}} & 0.7434\\ 
Anticoagulants / Antiplatelets (28,825; 45.75\%) & \textbf{\underline{0.7954}} & \textbf{0.7903}\\
Sedatives (26,562; 42.16\%) & \textbf{\underline{0.9182}} & \textbf{0.9149}\\
Analgesics (31,672; 50.27\%) & \textbf{\underline{0.8683}} & 0.8583\\
Neuromuscular blockers (2,995; 4.75\%) & \textbf{\underline{0.8889}} & \textbf{0.8831}\\
GI protection (21,096; 33.49\%) & \textbf{\underline{0.7160}} & \textbf{0.7060}\\
Blood products / Transfusions (9,205; 14.61\%) & \textbf{0.8995} & \textbf{\underline{0.9007}}\\
Parenteral nutrition (1,370; 2.17\%) & \textbf{\underline{0.8780}} & \textbf{0.8402}\\
\hline
\multicolumn{3}{l}{\textbf{\textit{Clinical Deterioration (adverse events within the next 24 hours)}}} \\
Severe hypoxemia (4,343; 6.89\%) & \textbf{\underline{0.7585}} & 0.6653\\
ECMO (1,369; 2.17\%) & \textbf{0.8486} & \textbf{\underline{0.8740}}\\
Invasive mechanical ventilation (27,257; 43.26\%) ${}^\ast$ & \textbf{\underline{0.9722}} & \textbf{0.9701}\\
Non-invasive mechanical ventilation (1,348; 2.14\%) & \textbf{\underline{0.9712}} & \textbf{0.9525}\\
Cardiac arrest (204; 0.32\%) & \textbf{\underline{0.9175}} & 0.8101\\
\hline
\multicolumn{3}{l}{\textbf{\textit{Organ dysfunction (SOFA subscore $\geq$ 2 within the next 24 hours)}}} \\
Respiratory (22,129; 85.33\%) & \textbf{\underline{0.7381}} & \textbf{0.7325}\\
Nervous system (28,824; 47.45\%) & \textbf{0.9346} & \textbf{\underline{0.9351}}\\
Cardiovascular (13,729; 21.79\%) & \textbf{\underline{0.8716}} & \textbf{0.8628}\\
Liver (5,144; 8.16\%) & \textbf{0.9324} & \textbf{\underline{0.9362}}\\
Coagulation (8,903; 14.13\%) & \textbf{\underline{0.9325}} & 0.9224\\
Kidneys (20,978; 33.30\%) ${}^\ast$ & \textbf{0.8457} & \textbf{\underline{0.8477}} \\
\hline
\textbf{Macro average} & \textbf{\underline{0.8650}} & 0.8583\\
\hline
\end{tabular}
\end{table*}

Table \ref{tab:performance} summarizes predictive performance across 33 ICU outcomes and identifies tasks for which integrating ECG waveforms provides a statistically significant advantage. Notably, there are no outcomes where the clinical tabular model is uniquely superior, as indicated by the absence of rows where only the unimodal model is highlighted. Instead, ECG waveform integration yields significant improvements in 14 of 33 outcomes, spanning mortality, medication administration, clinical deterioration, and organ dysfunction, and results in a significantly higher macro-average AUROC. Appendix~\ref{app:performance} presents additional results for a tree-based model as an alternative to the tabular-only model. This baseline achieves a lower macro-average AUROC (0.8460) compared to the RealMLP model, highlighting the superior performance of deep learning approaches over traditional tabular methods, in line with findings in \cite{erickson2025tabarena}.

\begin{figure*}[t]
    \centering
    \includegraphics[width=.8\textwidth]{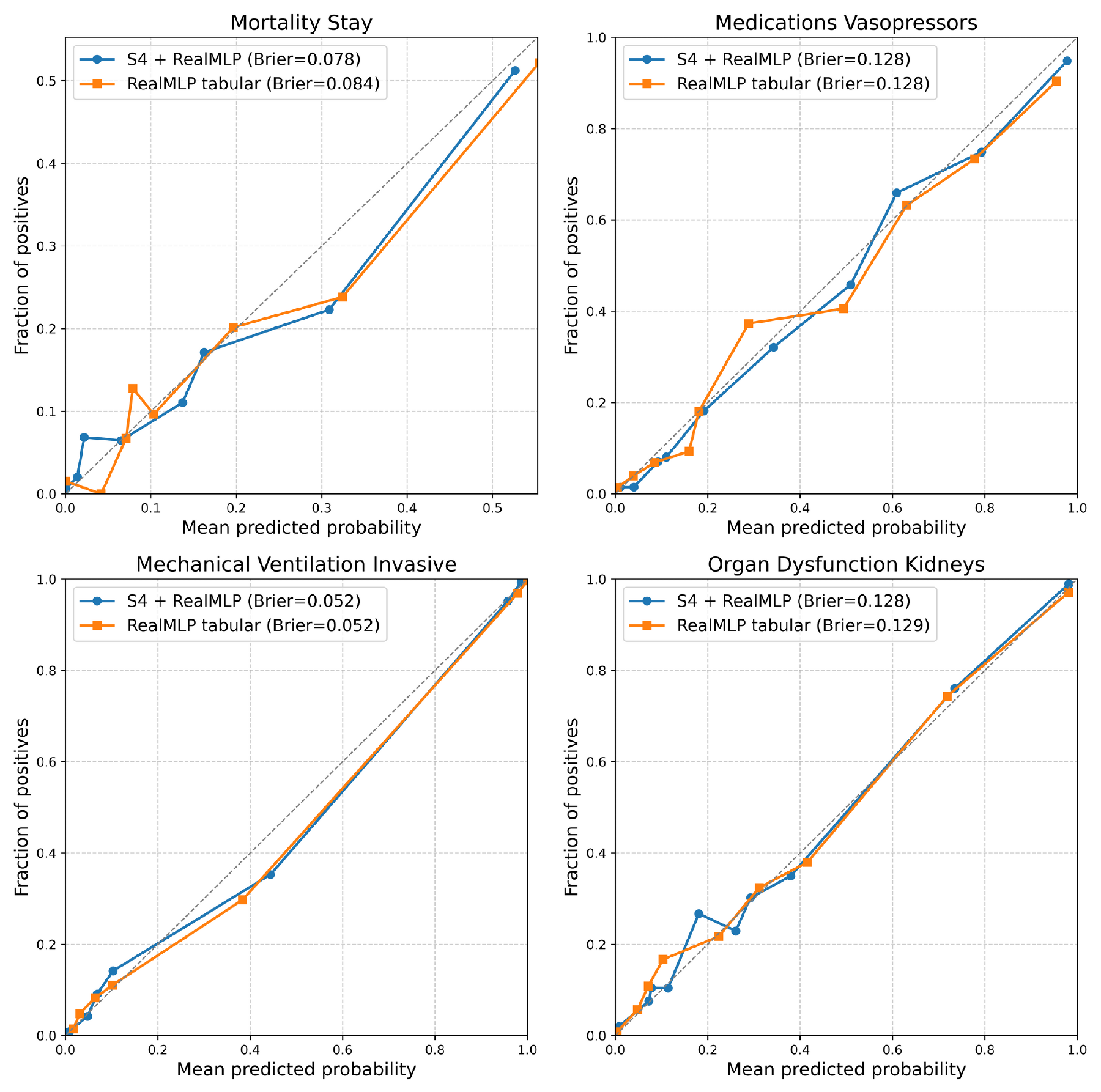} 
    \caption{Calibration plots for a subset of representative labels, one per category: mortality (within stay), invasive mechanical ventilation (clinical deterioration), vasopressors (medications), and SOFA kidney (organ dysfunction). Probabilities were calibrated using isotonic regression on the validation set. Both models are overall well-calibrated, with S4+RealMLP showing improved alignment with observed outcomes, particularly in the higher probability ranges. Calibration was computed using a quantile strategy with 10 bins.}
    \label{fig:calibration_main}
\end{figure*}

Performance gains are most evident for outcomes linked to acute physiological instability and cardiovascular dynamics. For mortality prediction, ECG-augmented models outperform tabular models for ICU mortality, stay mortality, and intermediate-term horizons (7–28 days), suggesting that short-term electrical signals provide prognostic information beyond static clinical variables. In medication prediction, waveform features significantly improve discrimination for antiarrhythmics, vasopressors, sedatives, analgesics, and parenteral nutrition—therapies typically initiated in response to evolving hemodynamic or electrophysiological changes.

The largest and most consistent benefits are observed in clinical deterioration tasks, where ECG waveforms substantially enhance prediction of severe hypoxemia, invasive and non-invasive mechanical ventilation, and cardiac arrest, reflecting their ability to capture early decompensation not evident in routine charted data. For organ dysfunction, waveform integration significantly improves prediction of cardiovascular and coagulation dysfunction, while remaining comparable for respiratory, neurological, hepatic, and renal outcomes. Overall, the improved macro-average AUROC indicates that multimodal integration provides a consistent benefit without degrading performance on any task.

\subsection*{Calibration}

Figure \ref{fig:calibration_main} shows that both S4+RealMLP and tabular-only RealMLP produce generally well-calibrated predictions across diverse clinical outcomes. Incorporating ECG waveforms (S4) leads to a measurable improvement: for mortality during the stay, the Brier score decreased from 0.084 to 0.078; for invasive mechanical ventilation, remained in 0.052; for vasopressors, remained in 0.128; and for SOFA kidney, from 0.129 to 0.128. This indicates that S4+RealMLP better aligns predicted probabilities with observed event frequencies, particularly in the higher probability ranges, reducing over- or underestimation. These results highlight that adding waveform-derived features improves not only discriminative performance but also the reliability of probabilistic predictions, which is crucial for informed clinical decision-making. See Appendix~\ref{app:calibration} for a complete set of figures across all investigated labels.

\begin{figure*}[t]
    \centering
    \includegraphics[width=.8\textwidth]{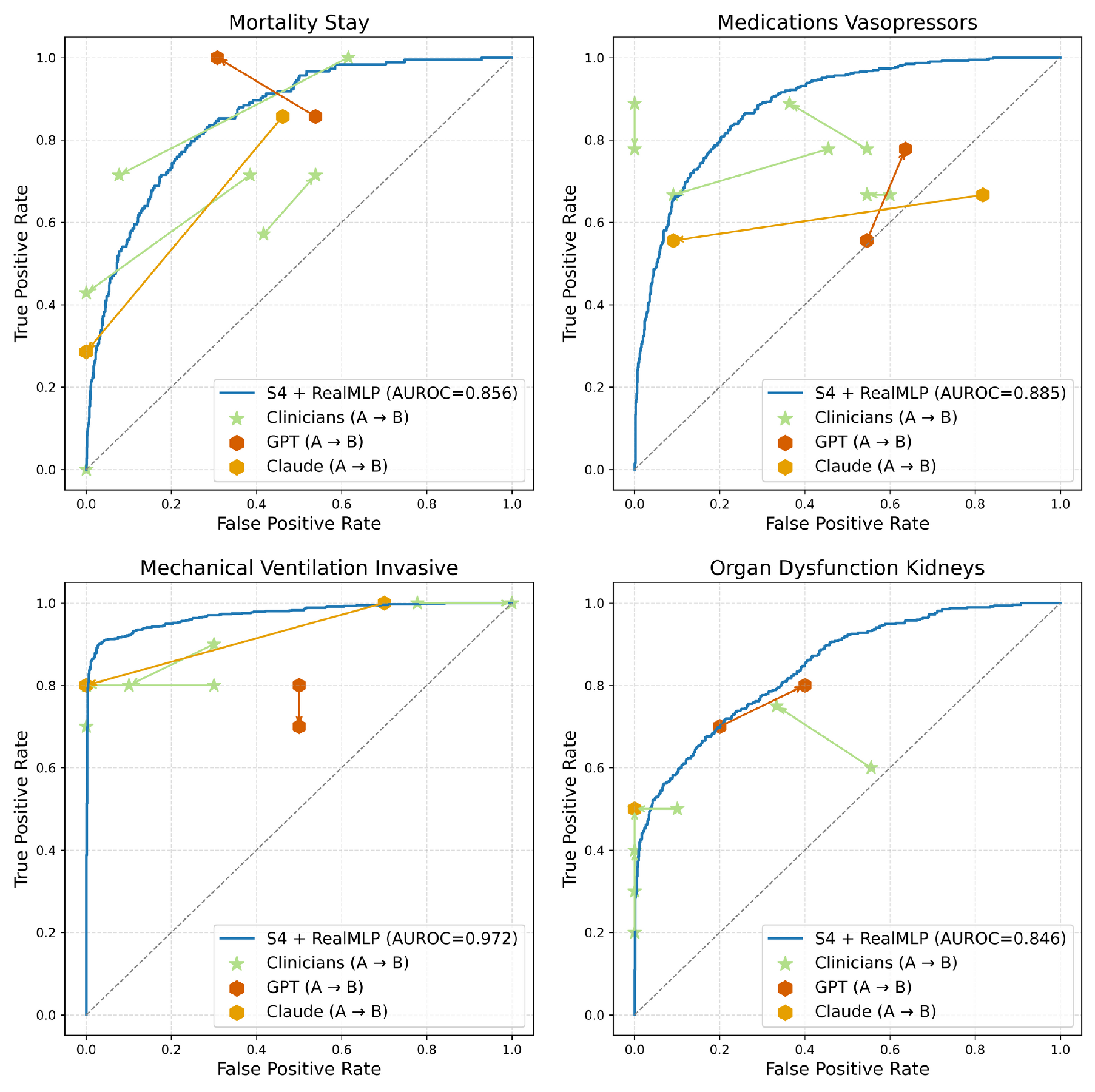} 
    \caption{Clinical benchmark plots for a subset of representative labels, one per category: mortality (within stay), invasive mechanical ventilation (clinical deterioration), vasopressors (medications), and SOFA kidney (organ dysfunction). Each plot contains the the main investigated model (S4+RealMLP) AUROC, as well as sensitivity and specificity of clinicians, GPT, and Claude for both benchmark A (clinician/LLM alone) and B (clinician/LLM+model).}
    \label{fig:clinical_benchmark}
\end{figure*}

\subsection*{Clinical benchmark}
\heading{Setup} To evaluate the added clinical value of our model, we selected a random subset of 20 test samples and shared them with 4 clinicians with relevant ICU expertise. For this benchmark, we focused on four key prediction targets, one from each category: stay mortality, vasopressors administration, invasive mechanical ventilation, and kidney dysfunction. In Benchmark A, clinicians were tasked to solve the four prediction tasks using the same information that was provided to the multimodal model. In Benchmark B, they were provided model predicted probabilities in order to refine their predictions.
Together, these two benchmarks provide a structured, reproducible assessment of the added value of multimodal AI predictions in ICU decision support. In addition, we include two proprietary LLMs (GPT 5.2 and Claude Sonnet 4.5) receiving only the clinical features as input. In order to avoid to specify a threshold when comparing continuous model predictions to binary clinicians' predictions and baselines, we plot these predictions along with the model's ROC curve, see Figure~\ref{fig:clinical_benchmark}.

\heading{Benchmark A: Model vs. clinicians}
Across the four tasks, clinicians exceed the model’s ROC performance in 12.5\% of the cases, perform comparably in 31.25\%, and underperform in 56.25\% of the cases, reflecting notable variability in their performance. LLMs exceed the model in 12.5\% of cases, are on par in 25\%, and underperform in 62.5\%, see Appendix~\ref{app:clinical benchmark} for details. These results indicate that, on average, the model outperforms both clinicians and LLMs for the considered tasks.

\heading{Benchmark B: Model-assisted decision-making}
The inclusion of the model predictions lead to notable improvement in model performance. Clinicians outperform the model in 37\% of the cases, perform on par in about 31\% of the cases and underperform only in about 31\% of the cases. Similarly, the augmented LLMs outperform in 37.5\% of the cases, perform on par in 25\% and underperform i 37.5\% of the cases. Beyond summary statistics, we consider differences in the Youden index (sensitivity + specificity – 1) to quantify improvement for each decision maker. Across all tasks, clinicians’ indices increased on average by 12\% when assisted by the model, while LLMs show a larger improvement of 16\%, see Appendix~\ref{app:clinical benchmark} for details. Overall, the model predictions enhance both clinicians and LLMs performance, by providing the greatest benefit to LLM-based decision-making for treatment-related tasks, while still offering substantial added value to clinicians on outcomes-related tasks.

\section*{Discussion}

\heading{Clinical significance} Table \ref{tab:performance} suggests that ECG waveforms provide complementary and clinically meaningful information beyond routine ICU data. The improvements in short- and medium-term mortality indicate that waveform dynamics capture acute physiological instability not fully reflected in standard clinical variables. Enhanced prediction of medications such as antiarrhythmics, sedatives, analgesics, vasopressors, and parenteral nutrition highlights the close coupling between cardiac electrophysiology and treatment intensity; in particular, these findings align with prior work demonstrating that ECG-based models can identify arrhythmic risk and guide antiarrhythmic therapy \cite{Ayyub2025AIarrhythmia,Siontis2021AIECGRisk}. ECG also improved detection of severe hypoxemia and cardiac arrest, emphasizing their sensitivity to acute physiological collapse. These results are consistent with evidence that hypoxemia induces detectable changes in heart rate variability, conduction, and repolarization \cite{Buchner2002HRVHypoxia,Somers1989HypoxiaSNS}, and that early electrical instability precedes cardiac arrest \cite{Attia2019LancetAIECG,Ahmad2017ECGArrest}. For organ dysfunction, waveform data were particularly informative for cardiovascular and coagulation outcomes, whereas other organ systems remained adequately modeled by clinical variables. Overall, integrating ECG waveforms into ICU decision support models could enhance early risk stratification, guide treatment prioritization \cite{koebe2025towards}, and improve monitoring of high-risk patients for deterioration events.

\heading{Calibration} Figure \ref{fig:calibration_main} illustrates that both S4+RealMLP and tabular-only RealMLP provide reasonably well-calibrated probability estimates across different clinical outcomes. Incorporating ECG waveforms (S4) generally improves calibration, particularly for higher predicted probabilities, reducing over- or underestimation of risk. This effect is most notable for ICU mortality and invasive mechanical ventilation, reflecting the additional signal captured by ECG features. Brier scores confirm this trend, with consistently lower values for the S4+RealMLP model. These results demonstrate that waveform-informed models not only improve discriminative performance but also provide more reliable probabilistic predictions, enhancing clinician trust and their utility for clinical decision support.

\heading{Clinical benchmark} The clinical benchmark reflected in Figure \ref{fig:clinical_benchmark} demonstrates that clinicians and LLMs on average underperform compared to the proposed model. However, when both of them have access to the model’s predictions, their performance improves consistently, achieving higher sensitivity and specificity across diverse tasks. This suggests that while the model provides strong standalone predictive power, its greatest value lies in supporting clinicians, enhancing their decision-making. The clinicians leveraging model predictions outperform/underperform the model in 37.5\%/31.25\% of the cases, respectively. If substantiated in larger user studies this would be an example of the most desirable augmentation scenario, where the augmented clinicians' performance exceeds the model performance \cite{vaccaro2024combinations}. At this point, one has to note that some tasks like medication administration prediction or the mechanical ventilation prediction represent therapy decisions and not predictions of the natural course of the patient trajectory, and might be affected by different clinical preferences.

\heading{Limitations}
While our dataset and methodology offer a significant step forward in multimodal ICU prediction, some limitations remain. Although our internal evaluation demonstrates strong performance, external validation could not be conducted due to the lack of publicly available datasets with comparable multimodal characteristics, particularly those including raw waveforms. We remain eager to apply our methods to external datasets as soon as suitable resources, even from private domains, become accessible. The set of input features is already very comprehensive, but does not cover current medication. While it was left out intentionally to avoid the complexity of different drugs and different dosages, it represents an important piece of information, which would typically be taken into account by an ICU clinician to assess a patient's state. The inclusion of medications as input features is a promising direction for future work.

\heading{Future work}
Future directions include the exploration of alternative tabular encoders and more sophisticated multimodal fusion strategies to better capture dependencies across data types. We also aim to investigate dynamic temporal modeling approaches that can handle irregularly sampled data natively, potentially reducing reliance on discrete sampling that often involves imputation and interval hyperparameter choice. Similarly, the inclusion of other waveforms and waveforms trends via embeddings.

To achieve real clinical impact and successful deployment, AI systems like the proposed must overcome critical barriers \cite{kelly2019key} including clinical integration challenges such as workflow adaptation and clinician trust, rigorous clinical validation ideally via randomized controlled trials, and regulatory frameworks that balance safety with innovation. Similarly, practical implementation also involves addressing data sharing and privacy and algorithm transparency \cite{he2019practical}. Effectively navigating these multifaceted challenges is essential for integrating AI into existing clinical workflows and meeting regulatory demands across diverse healthcare settings. In this context, foundational ECG models \cite{al2025benchmarking} provide a promising avenue, as pretrained ECG representations can enhance both discriminative performance and calibration across multiple clinical tasks.

Enhancing explainability before clinical adoption is an essential step to increase clinician trust and facilitate safe implementation in practice. Currently, methods such as those proposed in \cite{11272614} provide causal explanations for unimodal time series data, but have not been extended towards multimodal scenarios.

\section*{Methods}

\subsection*{Predictive workflow and dataset creation}

For each ECG recorded during an ICU stay, a single observation window is extracted from ICU admission up to the ECG timestamp, during which clinical features are collected. These inputs are used to train a unified model capable of addressing a broad set of clinically meaningful prediction tasks spanning the ICU stay and subsequent outcomes. The \mdsicu{} dataset integrates data from MIMIC-IV \cite{Johnson2023}, MIMIC-IV-ECG \cite{MIMICIVECG2023}, and MIMIC-IV-ECG-ICD \cite{strodthoff2024prospects}, focusing on ICU patients with at least one 10-second 12-lead ECG. ECG waveforms are included due to prior evidence demonstrating their statistically significant added value for related clinical prediction tasks \cite{alcaraz2024mds}. Data preprocessing involves rigorous feature cleaning, including outlier removal and correction of implausible values. The use of MIMIC is motivated by its breadth of patient populations and data modalities, providing a robust foundation for developing and benchmarking ICU decision-support systems. We highlight differences to existing ICU datasets and benchmarking frameworks in terms of source databases, populations, size, features, tasks, and open source availability in Appendix~\ref{app:icudatasets}.

\subsection*{Feature extraction}

We extracted a comprehensive set of clinical features aligned with each 10-second, 12-lead ECG to provide contextual information for predictive modeling. Each sample includes the raw ECG waveform capturing high-resolution cardiac electrical activity at the time of prediction, alongside demographic attributes (age, gender, ethnicity), biometric measurements (height, weight, body mass index), and vital signs summarized up to the ECG acquisition time. Vital signs span cardiovascular, respiratory, neurological, thermal, and oxygenation systems and reflect continuously monitored ICU physiology. Laboratory measurements provide quantitative insight into systemic physiology and organ function and include hematological, metabolic, electrolyte, inflammatory, perfusion, and cardiac markers, selected based on clinical relevance following \cite{moor2023predicting} to avoid label-driven bias. We further incorporate procedural context, including surgical interventions during the ICU stay or within the preceding 24 hours, as well as the use of supporting devices such as invasive and non-invasive mechanical ventilation within the patient context window. 

To model temporal dependencies in irregularly sampled vital signs and laboratory data, we apply a statistical feature extraction approach that aggregates descriptive statistics over the observation window. Features include the minimum, maximum, first and last values, and the time elapsed since the last observation. In addition, we compute a secondary time series from first-order differences and apply the same summarization to capture temporal variation. This dual-level encoding represents both absolute trends and dynamic changes, yielding a compact and expressive feature set well suited for temporal tabular clinical prediction models.

\subsection*{Prediction tasks}

Prediction tasks are grouped into four clinically relevant categories: mortalities, medications, clinical deterioration and organ dysfunction, covering both short- and long-term ICU decision-support needs. Mortality is predicted during the ICU and hospital stay at irregular intervals, as well as at fixed horizons of 1, 2, 7, 28, 90, 180, and 365 days from the prediction time. Medication prediction is formulated as binary classification of drug class administration within the next 24 hours, including crystalloids, electrolytes, antibiotics, vasopressors, inotropes, antiarrhythmics, anticoagulants/antiplatelets, sedatives, analgesics, neuromuscular blockers, gastrointestinal protection agents, blood products, and parenteral nutrition. Clinical deterioration targets acute adverse events within 24 hours, including severe hypoxemia, mechanical ventilation, in-hospital cardiac arrest, and extracorporeal membrane oxygenation (ECMO) initiation, supporting timely intervention and resource planning.
Organ dysfunction is assessed via early warning score prediction, specifically Sepsis-3 \cite{hofford2022opensep} defined by SOFA $\geq$ 2 \cite{lambden2019sofa}, over a 24-hour window to enable proactive risk stratification. Definitions for all prediction targets can be found in Appendix~\ref{app:targets}.

The final dataset consists of 33 labels across 63,001 samples from 27,062 patients across 33,590 ICU stays. Each input includes a single 10-second ECG waveform along with 801 tabular features.

\subsection*{Model architectures}

\begin{figure*}[!ht]
    \centering
    \includegraphics[width=\textwidth]{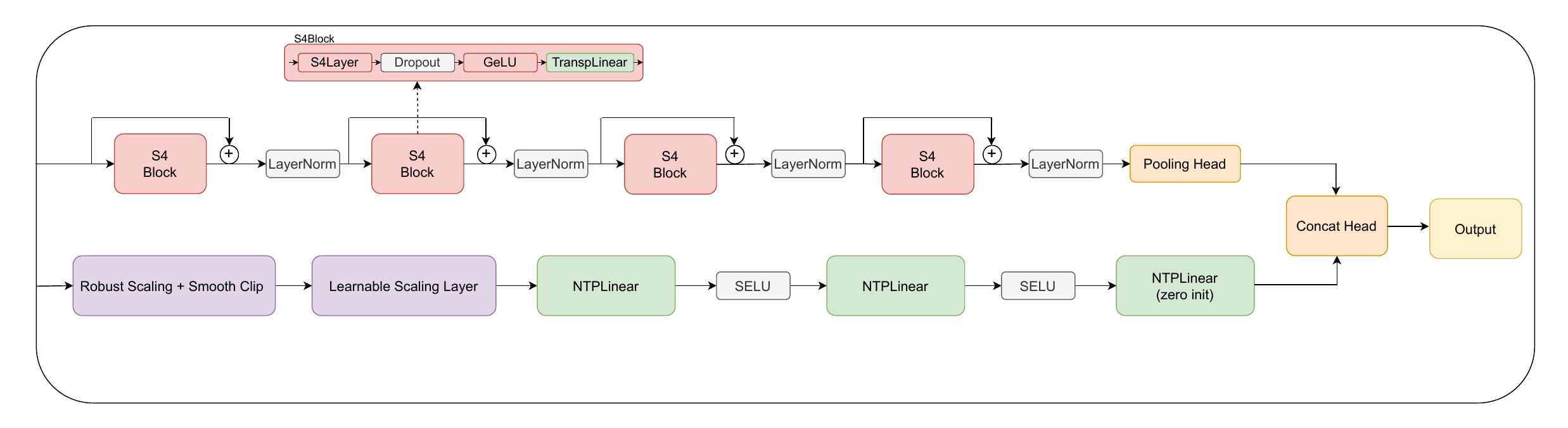}
    \caption{Overview of the multimodal architecture. The data flow proceeds from left to right. The top branch shows the time-series encoder based on structured state space models (S4), consisting of four S4 blocks followed by a pooling head. The bottom branch depicts the tabular encoder implemented using RealMLP, which applies robust quantile-based feature scaling, learnable feature re-scaling, and a stack of scaled linear layers (NTPLinear) with SELU activations. The outputs of both encoders are concatenated to form a joint multimodal representation used for prediction.}
    \label{fig:architectures}
\end{figure*}

We adopt a multimodal learning setup that combines ECG waveforms and clinical tabular features. ECG waveforms are processed using an S4-based encoder built on structured state space models \cite{Gu2021EfficientlyML}, which has demonstrated superior performance over conventional deep learning architectures such as CNNs, RNNs, and transformers in physiological time-series modeling \cite{strodthoff2024prospects, mehari2023towards}. The encoder consists of four stacked S4 blocks, each comprising an S4 layer, dropout, a GeLU activation function, and a transposed linear layer, with normalization layers applied between blocks. A pooling head is used to obtain a fixed-dimensional representation. Tabular clinical features are encoded using RealMLP \cite{holzmuller2024better}, an improved multilayer perceptron architecture that has been shown to perform competitively with state-of-the-art tree-based methods on tabular benchmarks \cite{erickson2025tabarena}. The RealMLP encoder applies a robust quantile-based scaling with smooth value clipping, followed by a learnable feature-scaling layer and three scaled linear layers (NTPLinear) with SELU activation functions. We employ a late fusion strategy by concatenating the representations produced by the two encoders to form a joint multimodal representation used for prediction. To assess the contribution of waveform data, we benchmark both the multimodal model and a tabular-only variant. Additionally, we report further tabular-only baselines using a tree-based model (XGBoost) in the appendix.

\subsection*{Training and evaluation}

We adopt stratified patient-wise data splits to ensure balanced representation across key clinical factors such as gender, age groups, and discharge diagnoses. Specifically, we divide the dataset into 20 stratified folds, allocating them to training, validation, and testing sets using a ratio of 18:1:1. The stratified folds are in line with the protocol from \cite{strodthoff2024prospects}. We apply median imputation based on training set statistics to handle missing data and introduce binary indicator columns to flag imputed values, following the protocol outlined in \cite{strodthoff2024prospects}, in line with best practices from the literature \cite{morvan2025imputation}.

We use AdamW \cite{loshchilov2018decoupled} as the optimizer with a learning rate of 0.001 and weight decay of 0.001, maintaining a constant learning rate schedule. Training is performed with a batch size of 64 for 20 epochs. Overfitting is controlled by selecting the best model based on validation performance. We minimize binary cross-entropy loss. Model performance is evaluated using macro average AUROC across all labels, and 95\% confidence intervals for AUROC are estimated using empirical bootstrap with 1000 iterations. The primary evaluation metric is macro AUROC, as it best reflects a model's discriminative power without requiring predefined decision thresholds, also in the presence of label imbalance \cite{mcdermott2024closer}. We assess the statistical significance of performance differences between two models by bootstrapping the performance difference between two models. The difference is considered statistically significant if the confidence interval does not cover the zero.

\subsection*{Clininical benchmark}
\heading{Clinicians' benchmark} For Benchmark A, clinicians received a single PDF file per patient sample containing all tabular features and a 12-lead ECG plot, and were asked to provide binary predictions for each of the four prediction targets. This allowed direct comparison between clinician judgment and model predictions. For Benchmark B, clinicians were given the same samples along with the model’s predicted probabilities, allowing them to revise their predictions based on the model output. This benchmark evaluated whether providing AI support improved clinician decision-making.

\heading{LLM baselines} We also assess the ability of proprietary LLMs to solve the four benchmarking tasks. For simplicity, we resorted to text-only LLMs, i.e. presented only the clinical data but not the ECG waveform. We used GPT 5.2 and Claude Sonnet 4.5 for this experiment. See Appendix~\ref{tab:youden_deterioration_transposed} for the specific prompt that was used to generate these both benchmarks.

\subsection*{Study protocol and reporting standards}

This study follows a comprehensive protocol aligned with established guidelines such as the Transparent Reporting of a Multivariable Prediction Model for Individual Prognosis or Diagnosis (TRIPOD). Detailed documentation is provided in the supplementary material.

\section*{Data Availability}
The datasets generated and/or analysed during this study are available from the publicly accessible MIMIC-IV, MIMIC-IV-ECG, and MIMIC-IV-ECG-ICD repositories, subject to the respective data use agreements. All preprocessing steps and code required to generate the dataset used in this study are provided in our open-access repository: \url{https://github.com/AI4HealthUOL/MDS-ICU}.
 
\section*{Code Availability}
All custom code used to generate the results reported in this manuscript, including preprocessing, feature extraction, and model training, is available in our repository: \url{https://github.com/AI4HealthUOL/MDS-ICU}. This code is sufficient to reproduce the analyses presented, and no restrictions apply beyond those specified by the underlying MIMIC-IV datasets.

\section*{Funding}
This work received no funding.

\section*{Author information}
Juan Miguel Lopez Alcaraz is the sole first author. Nils Strodthoff is the sole senior author.

\subsection*{Author contributions}

\textbf{Conceptualization:} JMLA, NS  
\textbf{Methodology:} JMLA, NS  
\textbf{Software:} JMLA, NS  
\textbf{Validation:} JMLA, XLM, EDZ, CH, WH, RK  
\textbf{Formal analysis:} JMLA  
\textbf{Investigation:} JMLA, XLM, EDZ, CH, WH, RK  
\textbf{Resources:} NS  
\textbf{Data curation:} JMLA  
\textbf{Writing – Original Draft:} JMLA, NS 
\textbf{Writing – Review \& Editing:} JMLA, NS, XLM, EDZ, CH, RK, WH  
\textbf{Visualization:} JMLA  
\textbf{Supervision:} NS  
\textbf{Project administration:} JMLA, NS

\subsection*{Corresponding author}
Correspondence should be addressed to \texttt{nils.strodthoff@uol.de}. ORCID \texttt{0000-0003-4447-0162}.

\section*{Ethics declaration}

\subsection*{Ethics approval}
The datasets used in this study are publicly available, de-identified, and have previously received ethical approval by their respective institutions. No additional ethical approval was required for this secondary data analysis.

\subsection*{Competing interests}
All authors declare no financial or non-financial competing interests.

\clearpage

\printbibliography

\appendix

\clearpage

\section{Predictive performance}
\label{app:performance}
Table~\ref{tab:performance_appendix} lists the predictive performance including confidence intervals via bootstrapping. In addition to the results from the main text, the table contains an additional tree-based baseline prediction model (XGBoost).

\begin{table*}[ht!]
\centering
\caption{Predictive performance (AUROC) comparison across models using ECG waveform–derived features and routine clinical data. Outcome counts and prevalence are reported in parentheses. Columns correspond to models trained with ECG waveforms plus RealMLP (S4+RealMLP), tabular-only RealMLP, and XGBoost. Columns ``Sig vs RealMLP'' and ``Sig vs XGBoost'' indicate paired bootstrap significance of S4+RealMLP. Confidence intervals represent 95\% CI from 1000 bootstrap iterations.}
\label{tab:performance_appendix}
\resizebox{\textwidth}{!}{
\begin{tabular}{lcccccc}
\hline
\textbf{Label (Counts, prevalence)} &
\textbf{S4+RealMLP} &
\textbf{RealMLP} &
\textbf{XGBoost} &
\textbf{Sig vs RealMLP} &
\textbf{Sig vs XGBoost} \\
\hline
\multicolumn{6}{l}{\textbf{\textit{Mortality}}} \\
ICU mortality (7,028; 11.16\%) & 0.8809 (0.8525--0.9082) & 0.8459 (0.8133--0.8809) & 0.8419 (0.8048--0.8754) & \checkmark & \checkmark \\
Stay mortality (9,443; 14.99\%) & 0.8561 (0.8301--0.8841) & 0.8356 (0.8052--0.8653) & 0.8283 (0.7974--0.8581) & \checkmark & \checkmark \\
1-day mortality (1,822; 2.89\%) & 0.9009 (0.8551--0.9435) & 0.8932 (0.8387--0.9416) & 0.8195 (0.7490--0.8864) &  & \checkmark \\
2-day mortality (2,744; 4.36\%) & 0.8834 (0.8375--0.9262) & 0.8809 (0.8345--0.9243) & 0.8425 (0.7890--0.8906) &  & \checkmark \\
7-day mortality (5,968; 9.47\%) & 0.8645 (0.8321--0.8956) & 0.8416 (0.8066--0.8766) & 0.8308 (0.7939--0.8653) & \checkmark & \checkmark \\
28-day mortality (11,128; 17.66\%) & 0.8609 (0.8372--0.8862) & 0.8349 (0.8092--0.8624) & 0.8459 (0.8212--0.8696) & \checkmark &  \\
90-day mortality (15,097; 23.96\%) & 0.8495 (0.8275--0.8715) & 0.8237 (0.7988--0.8469) & 0.8324 (0.8093--0.8561) & \checkmark & \checkmark \\
180-day mortality (17,681; 28.06\%) & 0.8393 (0.8188--0.8604) & 0.8166 (0.7935--0.8373) & 0.8298 (0.8088--0.8522) & \checkmark &  \\
1-year mortality (20,752; 32.94\%) & 0.8320 (0.8105--0.8531) & 0.8109 (0.7865--0.8324) & 0.8037 (0.7805--0.8253) & \checkmark & \checkmark \\
\hline
\multicolumn{6}{l}{\textbf{\textit{Medications}}} \\
Crystalloids (57,752; 91.67\%) & 0.8811 (0.8521--0.9077) & 0.8883 (0.8614--0.9129) & 0.8789 (0.8501--0.9073) &  &  \\
Electrolytes (33,568; 53.28\%) & 0.8017 (0.7755--0.9092) & 0.8101 (0.7995--0.9304) & 0.8101 (0.7960--0.9367) &  &  \\
Antibiotics (37,136; 58.95\%) & 0.8570 (0.8285--0.9092) & 0.8562 (0.7995--0.9304) & 0.8562 (0.7961--0.9367) &  &  \\
Vasopressors (21,012; 33.35\%) & 0.8854 (0.8575--0.9092) & 0.8781 (0.7995--0.9304) & 0.8781 (0.7961--0.9367) &  &  \\
Inotropes (5,746; 9.12\%) & 0.8732 (0.8471--0.9092) & 0.8738 (0.7995--0.9304) & 0.8738 (0.7961--0.9367) &  &  \\
Antiarrhythmics (6,663; 10.58\%) & 0.7909 (0.7614--0.8283) & 0.7434 (0.7995--0.9304) & 0.7434 (0.7961--0.9367) & \checkmark &  \\
Anticoagulants / Antiplatelets (28,825; 45.75\%) & 0.7954 (0.7755--0.8283) & 0.7903 (0.7995--0.9304) & 0.7903 (0.7961--0.9367) & \checkmark &  \\
Sedatives (26,562; 42.16\%) & 0.9182 (0.8931--0.9416) & 0.9149 (0.8703--0.9187) & 0.9149 (0.8362--0.9367) & \checkmark &  \\
Analgesics (31,672; 50.27\%) & 0.8683 (0.8483--0.8918) & 0.8583 (0.7995--0.9304) & 0.8583 (0.7961--0.9367) & \checkmark &  \\
Neuromuscular blockers (2,995; 4.75\%) & 0.8889 (0.8591--0.9092) & 0.8831 (0.7995--0.9304) & 0.8831 (0.7961--0.9367) & \checkmark &  \\
GI protection (21,096; 33.49\%) & 0.7160 (0.6896--0.7420) & 0.7060 (0.6752--0.7323) & 0.7060 (0.6553--0.7317) & \checkmark &  \\
Blood products / Transfusions (9,205; 14.61\%) & 0.8995 (0.8732--0.9007) & 0.9007 (0.7995--0.9304) & 0.9007 (0.7961--0.9367) &  &  \\
Parenteral nutrition (1,370; 2.17\%) & 0.8780 (0.8402--0.9092) & 0.8402 (0.7995--0.9304) & 0.8402 (0.7961--0.9367) & \checkmark &  \\
\hline
\multicolumn{6}{l}{\textbf{\textit{Clinical Deterioration}}} \\
Severe hypoxemia (4,343; 6.89\%) & 0.7585 (0.7083--0.8088) & 0.6653 (0.6023--0.7179) & 0.6653 (0.6023--0.7179) & \checkmark & \checkmark \\
ECMO (1,369; 2.17\%) & 0.8486 (0.8207--0.8740) & 0.8740 (0.7995--0.9304) & 0.8740 (0.7961--0.9367) &  &  \\
Invasive mechanical ventilation (27,257; 43.26\%) & 0.9722 (0.9639--0.9795) & 0.9701 (0.9623--0.9783) & 0.9701 (0.9613--0.9786) & \checkmark &  \\
Non-invasive mechanical ventilation (1,348; 2.14\%) & 0.9712 (0.9455--0.9904) & 0.9525 (0.9380--0.9914) & 0.9525 (0.9148--0.9817) & \checkmark &  \\
Cardiac arrest (204; 0.32\%) & 0.9175 (0.8279--1.0000) & 0.8101 (0.7432--1.0000) & 0.8101 (0.6978--0.9458) & \checkmark & \checkmark \\
\hline
\multicolumn{6}{l}{\textbf{\textit{Organ dysfunction}}} \\
Respiratory (22,129; 85.33\%) & 0.7381 (0.7083--0.8088) & 0.7325 (0.6964--0.7987) & 0.7325 (0.6023--0.7179) & \checkmark & \checkmark \\
Nervous system (28,824; 47.45\%) & 0.9346 (0.9229--0.9459) & 0.9351 (0.9196--0.9433) & 0.9351 (0.9232--0.9466) &  &  \\
Cardiovascular (13,729; 21.79\%) & 0.8716 (0.8279--0.9092) & 0.8628 (0.7995--0.9304) & 0.8628 (0.7961--0.9367) & \checkmark &  \\
Liver (5,144; 8.16\%) & 0.9324 (0.9074--0.9542) & 0.9362 (0.8724--0.9397) & 0.9362 (0.9079--0.9590) &  &  \\
Coagulation (8,903; 14.13\%) & 0.9325 (0.8279--0.9542) & 0.9224 (0.7433--0.9314) & 0.9224 (0.6990--0.9458) & \checkmark & \checkmark \\
Kidneys (20,978; 33.30\%) & 0.8457 (0.8083--0.8680) & 0.8477 (0.8316--0.8719) & 0.8477 (0.8272--0.8686) &  &  \\
\hline
\textbf{Macro} & 0.8650 (0.8561--0.8733) & 0.8583 (0.8493--0.8676) & 0.8460 (0.8362--0.8554) & \checkmark & \checkmark \\
\hline
\end{tabular}
}
\end{table*}

\clearpage
\section{Calibration}
\label{app:calibration}
\begin{figure*}[!ht]
    \centering
    \includegraphics[width=\textwidth]{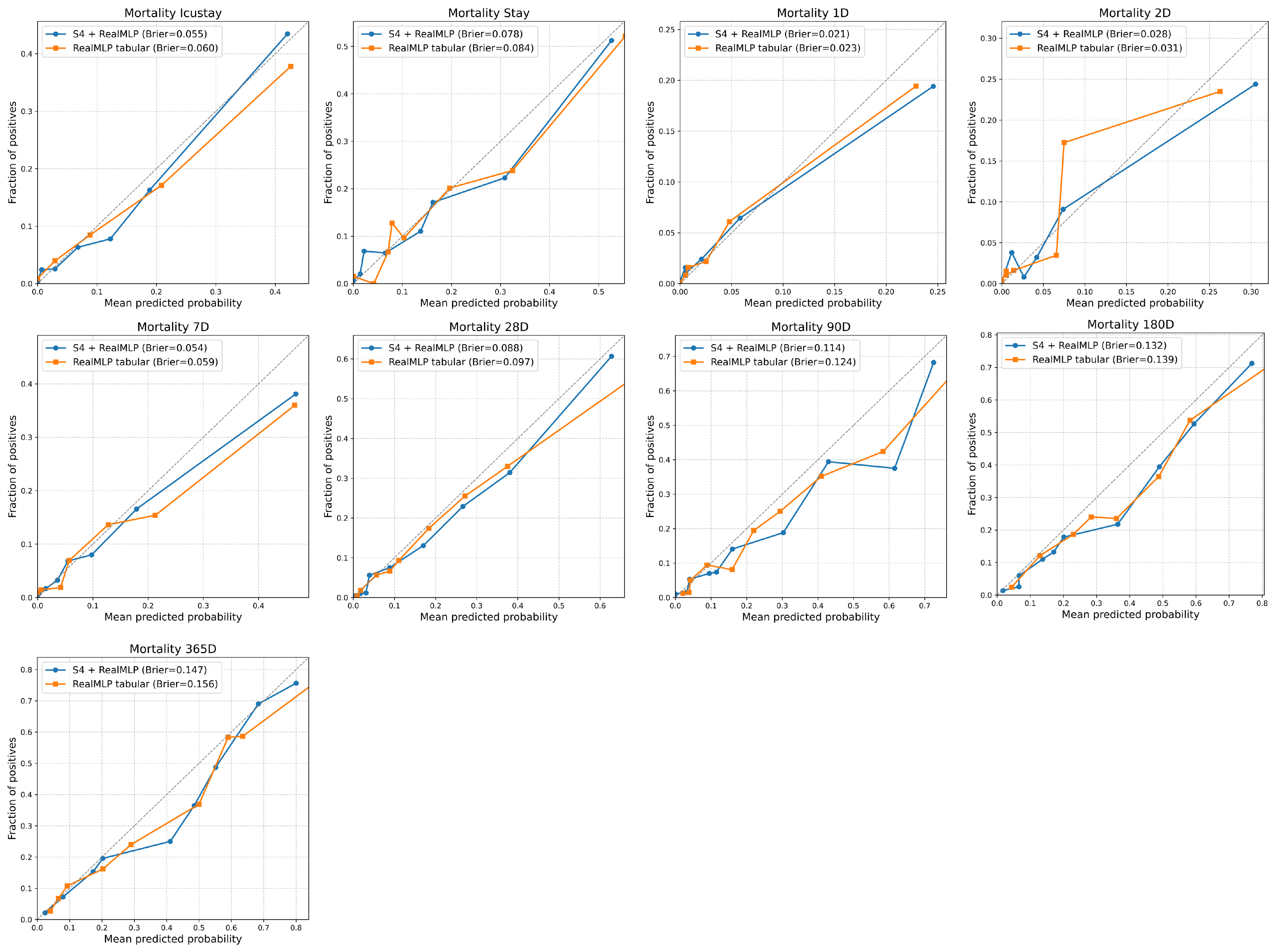} 
    \caption{Calibration figures for the mortality category.}
    \label{fig:calibration_mortality}
\end{figure*}

\begin{figure*}[!ht]
    \centering
    \includegraphics[width=\textwidth]{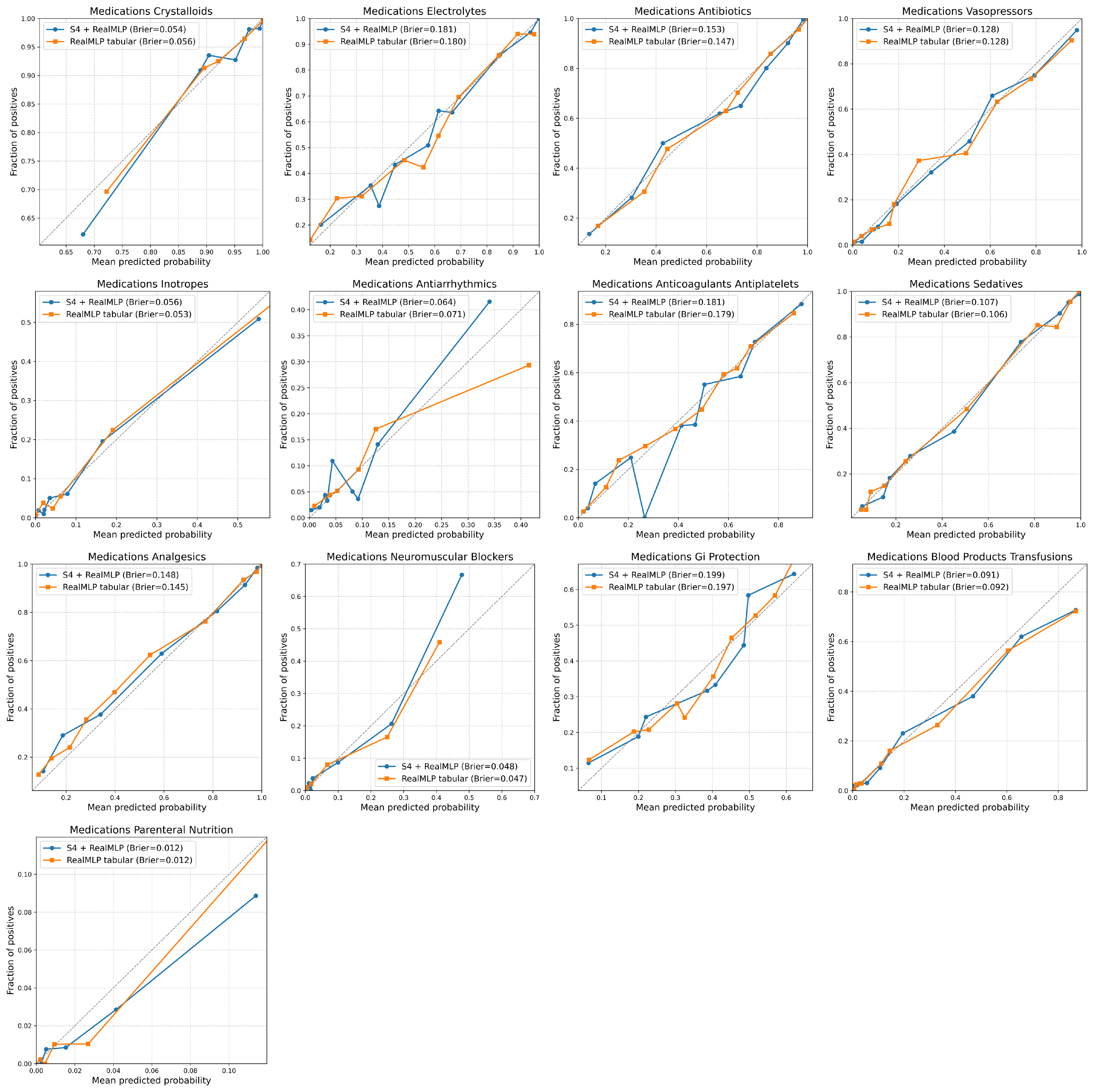} 
    \caption{Calibration figures for the medications category.}
    \label{fig:calibration_medications}
\end{figure*}

\begin{figure*}[!ht]
    \centering
    \includegraphics[width=\textwidth]{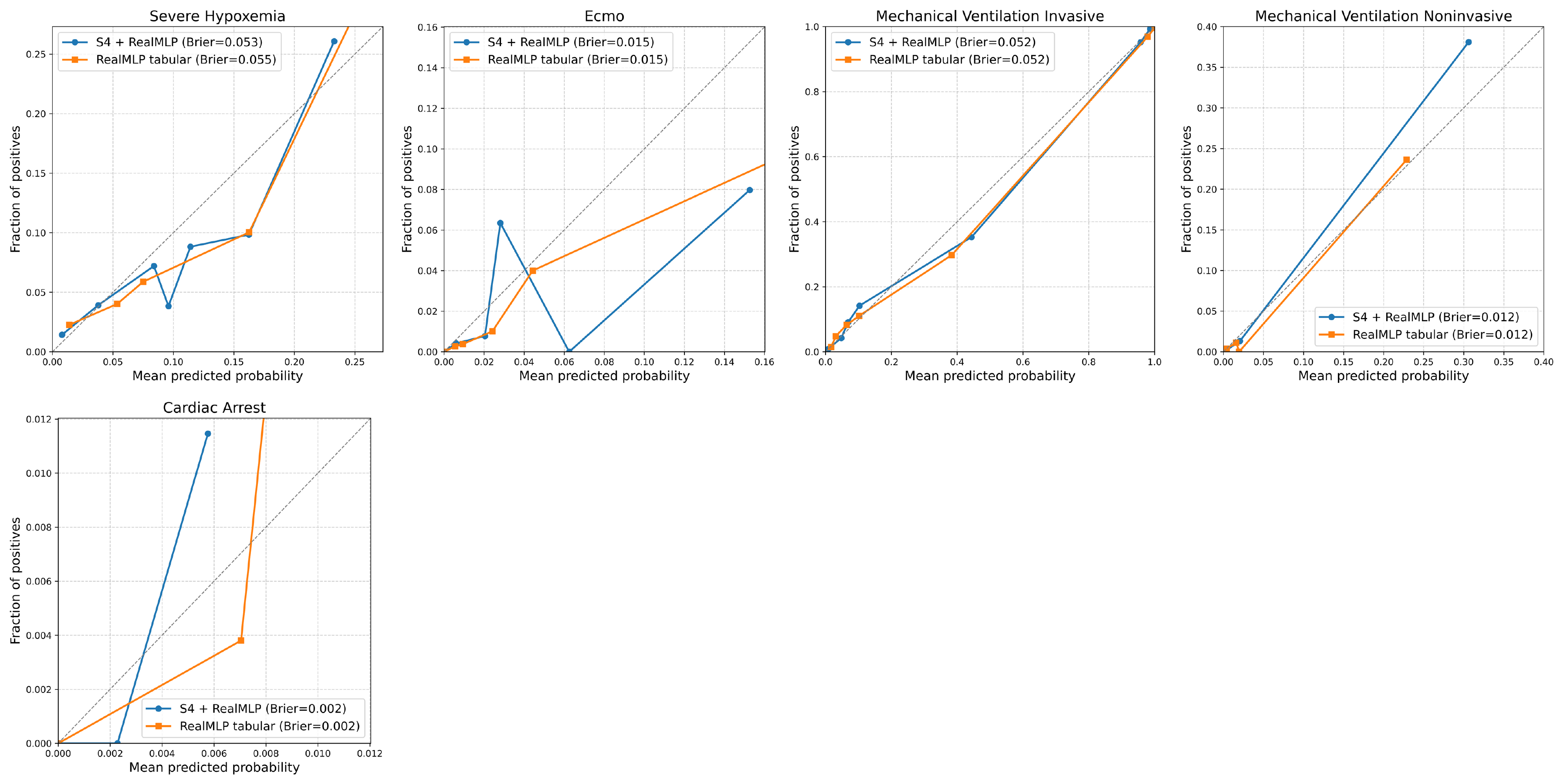} 
    \caption{Calibration figures for the clinical deterioration category.}
    \label{fig:calibration_deterioration}
\end{figure*}

\begin{figure*}[!ht]
    \centering
    \includegraphics[width=\textwidth]{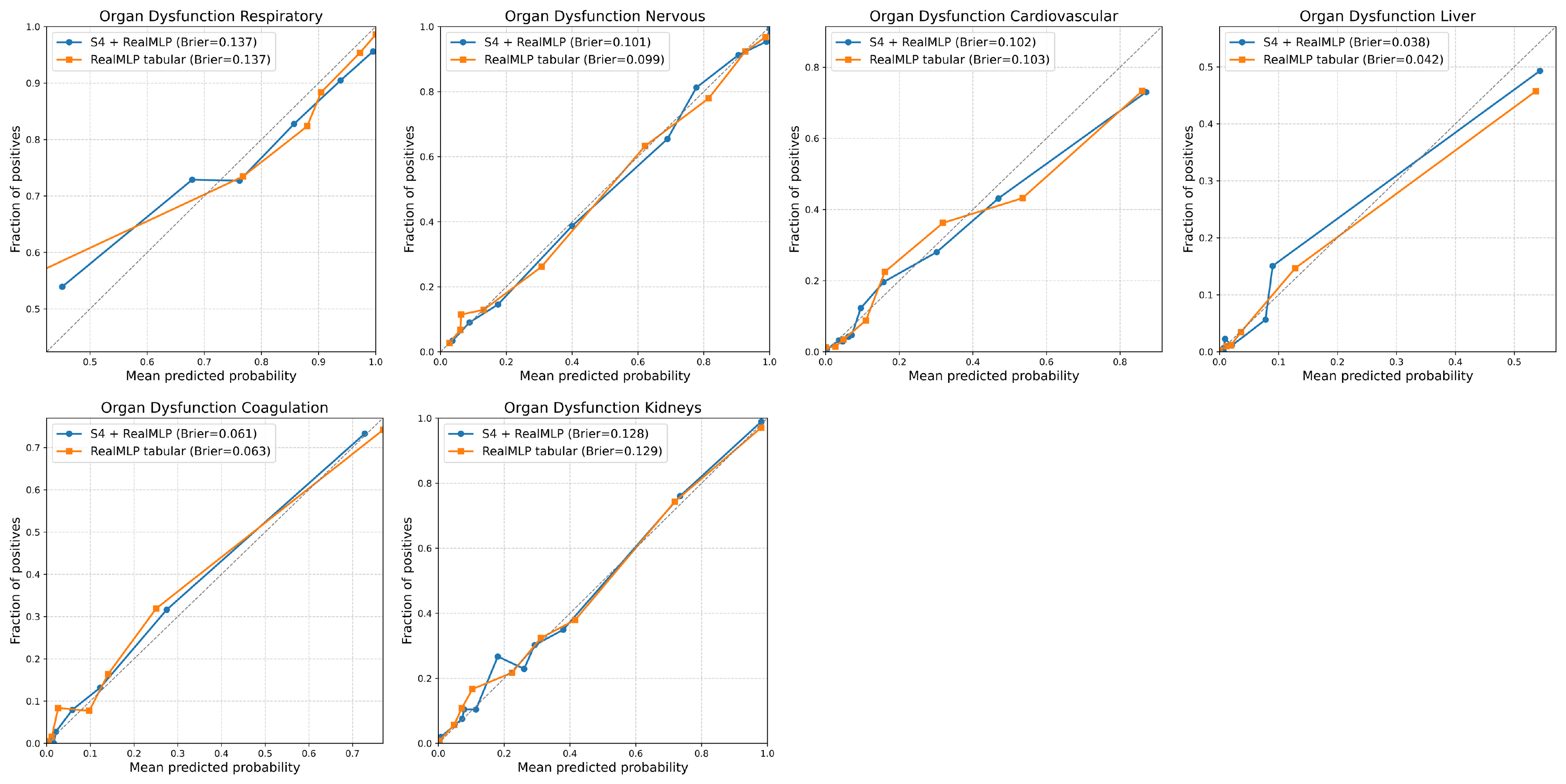} 
    \caption{Calibration figures for the organ dysfunction category.}
    \label{fig:calibration_sofa}
\end{figure*}

Tables~\ref{fig:calibration_mortality}-\ref{fig:calibration_sofa} shows calibration curves for all considered prediction targets.

\clearpage

\section{Clinical benchmark: Additional results}
\label{app:clinical benchmark}

Table \ref{tab:roc_summary1} and \ref{tab:roc_summary2} reports the counts below, comparable, or above the model for clinicians and LLMs across all tasks for both benchmarks. Table \ref{tab:youden_deterioration_transposed} reports the Youden index for four clinicians, two proprietary LLMs (GPT 5.2 and Claude Sonnet 4.5), and the baseline model across four clinical deterioration labels. Values for benchmark A → B and their corresponding changes (Δ) are shown for each model.

\begin{table}[ht]
\centering
\caption{Summary of decision makers' performance relative to the model's ROC curve for benchmark A: model vs clinicians. Counts indicate the number of instances above, comparable, or below the model (by comparing to the model's ROC curve). Percentages in the last row are computed relative to total observations for clinicians (16) and LLMs (8).}
\label{tab:roc_summary1}
\begin{tabular}{lcc}
\hline
\textbf{Task} & \textbf{Clinicians} & \textbf{LLMs} \\
\hline
Mortality              & 1 / 1 / 2       & 0 / 0 / 2       \\
Vasopressors           & 1 / 0 / 3       & 0 / 0 / 2        \\
Mechanical ventilation & 0 / 2 / 2       & 0 / 1 / 1    \\
Kidney dysfunction     & 0 / 2 / 2       & 1 / 1 / 0      \\
\hline
Sum               & 2 / 5 / 9    & 1 / 2 / 5  \\
Sum (\%)               & 12.5 / 31.25 / 56.25    & 12.5 / 25 / 62.5  \\
\hline
\end{tabular}
\end{table}

\begin{table}[ht]
\centering
\caption{Summary of decision makers' performance relative to the model's ROC curve for benchmark B: model and clinicians. Counts indicate the number of instances above, comparable, or below the model (by comparing to the model's ROC curve). Percentages in the last row are computed relative to total observations for clinicians (16) and LLMs (8).}
\label{tab:roc_summary2}
\begin{tabular}{lcc}
\hline
\textbf{Task} & \textbf{Clinicians} & \textbf{LLMs} \\
\hline
Mortality              & 3 / 0 / 1       & 2 / 0 / 0       \\
Vasopressors           & 1 / 1 / 2       & 0 / 1 / 1        \\
Mechanical ventilation & 0 / 3 / 1       & 0 / 1 / 1    \\
Kidney dysfunction     & 2 / 1 / 1       & 1 / 0 / 1      \\
\hline
Sum               & 6 / 5 / 5    & 3 / 2 / 3  \\
Sum (\%)               & 37.5 / 31.25 / 31.25    & 37.5 / 25 / 37.5  \\
\hline
\end{tabular}
\end{table}

\begin{table*}[ht!]
\centering
\caption{Youden index comparison across clinicians, and LLMs for different clinical deterioration labels.}
\label{tab:youden_deterioration_transposed}
\resizebox{\textwidth}{!}{
\begin{tabular}{lcccc}
\hline
\textbf{Model / Label} & \textbf{Mortality Stay} & \textbf{Medications Vasopressors} & \textbf{Mechanical Ventilation Invasive} & \textbf{Organ Dysfunction Kidneys} \\
\hline
Clinician 1 & 0.330 → 0.429 (Δ=0.099) & 0.232 → 0.525 (Δ=0.293) & 0.600 → 0.700 (Δ=0.100) & 0.200 → 0.500 (Δ=0.300) \\
Clinician 2 & 0.385 → 0.637 (Δ=0.253) & 0.323 → 0.576 (Δ=0.253) & 0.500 → 0.800 (Δ=0.300) & 0.400 → 0.500 (Δ=0.100) \\
Clinician 3 & 0.000 → 0.000 (Δ=0.000) & 0.889 → 0.778 (Δ=-0.111) & 0.700 → 0.700 (Δ=0.000) & 0.300 → 0.400 (Δ=0.100) \\
Clinician 4 & 0.155 → 0.176 (Δ=0.021) & 0.067 → 0.121 (Δ=0.055) & 0.222 → 0.000 (Δ=-0.222) & 0.044 → 0.417 (Δ=0.372) \\
GPT 5.2 & 0.319 → 0.692 (Δ=0.374) & 0.010 → 0.141 (Δ=0.131) & 0.300 → 0.200 (Δ=-0.100) & 0.500 → 0.400 (Δ=-0.100) \\
Claude Sonnet 4.5 & 0.396 → 0.286 (Δ=-0.110) & -0.152 → 0.465 (Δ=0.616) & 0.300 → 0.800 (Δ=0.500) & 0.500 → 0.500 (Δ=0.000) \\
\hline
\end{tabular}
}
\end{table*}

\clearpage

\section{Definition of prediction targets}
\label{app:targets}
Tables~\ref{tab:medication}-\ref{tab:clinical_deterioration} provide specific definitions for the prediction targets considered in this work.

\begin{table}[!ht]
\centering
\caption{Descriptive table defining medication label groups based on specific drugs. The left column lists medication labels, and the right column details the drugs tracked. A positive label is assigned if any drug from the corresponding group was administered within the evaluated time horizon}
\label{tab:medication}
\small
\begin{tabular}{l p{10cm}}
\textbf{Label} & \textbf{Drugs} \\
\hline
medication\_crystalloids & NaCl 0.9\%, Dextrose 5\%, Free Water, LR, D5NS, D5 1/2NS, D5LR, D5 1/4NS, NaCl 0.45\%, Sterile Water, Dextrose 10\%, Dextrose 20\%, Dextrose 30\%, Dextrose 40\%, Dextrose 50\%, NaCl 3\% (Hypertonic Saline), NaCl 23.4\% \\
\hline
medication\_electrolytes & Potassium Chloride, KCL (Bolus), KCl (CRRT), K Phos, Na Phos, Calcium Gluconate, Calcium Gluconate (CRRT), Calcium Gluconate (Bolus), Calcium Chloride, Magnesium Sulfate, Magnesium Sulfate (Bolus), Magnesium Sulfate (OB-GYN), Sodium Bicarbonate 8.4\%, Sodium Bicarbonate 8.4\% (Amp), Hydrochloric Acid - HCL \\
\hline
medication\_antibiotics & Cefepime, Vancomycin, Ceftriaxone, Levofloxacin, Azithromycin, Metronidazole, Bactrim (SMX/TMP), Cefazolin, Ciprofloxacin, Meropenem, Piperacillin/Tazobactam (Zosyn), Piperacillin, Omeprazole (Prilosec), Tobramycin, Doxycycline, Linezolid, Daptomycin, Ceftazidime, Ampicillin/Sulbactam (Unasyn), Ampicillin, Acyclovir, Clindamycin, Aztreonam, Colistin, Amikacin, Imipenem/Cilastatin, Ceftaroline, Rifampin, Erythromycin, Gentamicin, Nafcillin, Tamiflu, Penicillin G potassium, Keflex, Quinine, Isoniazid, Ethambutol, Pyrazinamide \\
\hline
medication\_vasopressors & Epinephrine, Epinephrine., Norepinephrine, Vasopressin, Dobutamine, Dopamine, Phenylephrine, Phenylephrine (50/250), Phenylephrine (200/250), Isuprel, Angiotensin II (Giapreza) \\
\hline
medication\_inotropes & Epinephrine, Epinephrine., Dobutamine, Dopamine, Isuprel, Milrinone \\
\hline
medication\_antiarrhythmics & Amiodarone, Amiodarone 600/500, Amiodarone 450/250, Amiodarone 150/100, Esmolol, Lidocaine, Procainamide, Verapamil, Diltiazem, Adenosine \\
\hline
medication\_anticoagulants\_antiplatelets & Heparin Sodium, Heparin Sodium (Prophylaxis), Heparin Sodium (Impella), Heparin Sodium (CRRT-Prefilter), Enoxaparin (Lovenox), Bivalirudin (Angiomax), Eptifibatide (Integrilin), Coumadin (Warfarin), Argatroban, Fondaparinux, Tirofiban (Aggrastat), Abciximab (Reopro), Lepirudin, ACD-A Citrate (1000ml), ACD-A Citrate (500ml), Citrate, Protamine sulfate \\
\hline
medication\_sedatives & Propofol, Midazolam (Versed), Lorazepam (Ativan), Diazepam (Valium), Dexmedetomidine (Precedex), Ketamine, Pentobarbital \\
\hline
medication\_analgesics & Fentanyl, Fentanyl (Concentrate), Morphine Sulfate, Hydromorphone (Dilaudid), Meperidine (Demerol), Acetaminophen-IV, Methadone Hydrochloride, Ketorolac (Toradol), Naloxone (Narcan) \\
\hline
medication\_neuromuscular\_blockers & Vecuronium, Rocuronium, Cisatracurium, Neostigmine (Prostigmin) \\
\hline
medication\_gi\_protection & Ranitidine (Prophylaxis), Pantoprazole (Protonix), Famotidine (Pepcid), Lansoprazole (Prevacid), Omeprazole (Prilosec), Carafate (Sucralfate), Esomeprazole (Nexium), Ranitidine \\
\hline
medication\_blood\_products\_transfusions & Packed Red Blood Cells (pRBCs), Platelets, Fresh Frozen Plasma (FFP), Cryoprecipitate, Whole Blood, Albumin 5\%, Albumin 25\%, IVIG (Intravenous Immunoglobulin), Factor VIII, Factor IX, Prothrombin Complex Concentrate (PCC), Recombinant Factor VIIa, Fibrinogen Concentrate, Thrombin, Tranexamic Acid (TXA), Erythropoietin (EPO), Iron Sucrose (Venofer), Iron Dextran, Iron Gluconate, Ferumoxytol \\
\hline
medication\_parenteral\_nutrition & TPN w/ Lipids, TPN without Lipids, Peripheral Parenteral Nutrition, Dextrose PN, Amino Acids, Lipids 20\%, Lipids 10\%, Lipids (additive) \\
\hline
\end{tabular}
\end{table}

\begin{table}[ht]
\centering
\caption{Descriptive table defining early warning score labels based on specific SOFA score (Sepsis-3) definition. The left column list the early warning score based od diverse physiological systems and the right columns details the criteria for its definition. A positive label is assigned if an event happened within the next 24 hours}
\label{tab:sofa}
\begin{tabular}{l p{10cm}}
\toprule
\textbf{Label} & \textbf{Definition} \\
\midrule
sofa\_respiratory & PaO$_2$/FiO$_2$ (mmHg) $< 300$ \\
sofa\_nervous & GCS $\leq 12$ \\
sofa\_cardiovascular & Vasopressor administration (e.g., dopamine, dobutamine, epinephrine, norepinephrine) \\
sofa\_liver & Bilirubin (mg/dl) $\geq 2$ \\
sofa\_coagulation & Platelets ($\times 10^3$/ml) $< 100$ \\
sofa\_kidneys & Creatinine (mg/dl) $\geq 2$ or urine output $< 500$ ml/day \\
\bottomrule
\end{tabular}
\end{table}

\begin{table}[ht]
\centering
\caption{Descriptive table defining clinical deterioration labels based on specific events. The left column list the clinical deterioration labels, and the right columns details the events tracked. A positive label is assigned if any event from the corresponding group happened within the valuation time horizon}
\label{tab:clinical_deterioration}
\begin{tabular}{l p{10cm}}
\hline
\textbf{Label} & \textbf{Definition} \\
\hline
severe\_hypoxemia &  O2 saturation pulseoxymetry  $<=$ 85 \\ \hline
ecmo &  Procedure: Sheath (Venous). Chartevents: Circuit Configuration (ECMO), Speed (ECMO), Flow (ECMO), Sweep (ECMO), Flow Alarm (Lo) (ECMO), Flow Alarm (Hi) (ECMO), FiO2 (ECMO), Suction events (ECMO), Cannula sites visually inspected (ECMO), Oxygenator visible (ECMO), Pump plugged into RED outlet (ECMO), Circuit inspected for clot (ECMO), P1 - P2 (ECMO), P1 (ECMO), P2 (ECMO), Emergency Equipment at bedside (ECMO), Flow Sensor repositioned (ECMO), Oxygenator/ECMO, ECMO \\ \hline
ihca &  Procedure: Cardioversion/Defibrillation \\ \hline
mechanical\_ventilation\_invasive &  Procedures: Invasive Ventilation \\ \hline
mechanical\_ventilation\_invasive &  Procedures: Non-invasive Ventilation \\ \hline
\end{tabular}
\end{table}

\clearpage

\section{Related ICU datasets}
\label{app:icudatasets}
Existing ICU datasets vary in scope and modality. EHRSHOT \cite{wornow2024ehrshot} provides longitudinal medication data across 15 tasks but lacks biometrics, waveforms, and vital trends. The work in \cite{sheikhalishahi2020benchmarking} use eICU with demographics, vitals, and labs across 770 tasks but omit temporal trends and waveforms. HiRID \cite{yeche2021hirid} offers detailed ICU stay data over 23 labels but excludes physiological waveforms. YAIB \cite{water2024yet} aggregates multiple sources for a large, diverse cohort but provides only five tasks and no waveform or trend data. Gupta et al. \cite{gupta2022extensive} uses MIMIC-IV with broad demographic, vital, and lab coverage over nine tasks but without waveforms. In contrast, \mdsicu{} builds on MIMIC-IV to integrate waveform signals and temporal trends of vitals and labs, supporting a large number of prediction tasks with multimodal, temporally rich representations, uniquely suited for advanced ICU time series modeling. Furthermore, we incorporate a benchmark with relevant medical specialists to further validate clinical utility. Relevant related datasets are compared in Table~\ref{tab:comparison}.

\begin{table*}[ht!]
\caption{Direct comparison with related works in terms of dataset size, features, availability, and number of target labels.}
\label{tab:comparison}
\centering
\scriptsize
\begin{tabular}{l l c c c c c c}
\toprule
\textbf{Name} & \textbf{Detail} & \textbf{EHRSHOT} & \textbf{Sheik. et al} & \textbf{Yeche et al} & \textbf{YAIB} & \textbf{Gupta et al}  & \textbf{\mdsicu{}} \\ 
\midrule
Reference & & \cite{wornow2024ehrshot} & \cite{sheikhalishahi2020benchmarking} & \cite{yeche2021hirid} & \cite{water2024yet} & \cite{gupta2022extensive}  & This work \\
\midrule
DB source & & Own & eICU & HiRID & Multiple & MIMIC-IV & MIMIC-IV \\ 
\midrule
Population & & Longitudinal & ICU & ICU & ICU & Longitudinal  & ICU \\
\midrule
\multirow{2}{*}{Size} & Patients & 6739 & 52325 & -$^{1}$ & +233696$^{1}$ & 65366 & 27,062 \\
& Visits & 921499 & 73718 & 33905 & 313400 & 94458 & 33,590 \\

\midrule
\multirow{7}{*}{Features} & Demographics & \ding{51} & \ding{51} & \ding{51} & \ding{51} & \ding{51} &  \ding{51} \\
& Biometrics & \ding{55} & \ding{51} & \ding{51} & \ding{51} & \ding{51}&  \ding{51} \\
& Vital parameters & \ding{55} & \ding{51} & \ding{51} & \ding{51}  &\ding{51} &  \ding{51}($\mathbb{T}$) \\
& Lab. values & \ding{51} & \ding{51} & \ding{51} & \ding{51} & \ding{51} & \ding{51}($\mathbb{T}$) \\
& Waveforms & \ding{55} & \ding{55} & \ding{55} & \ding{55} & \ding{55} & \ding{51} \\
& Chief complaint & \ding{55} & \ding{55} & \ding{55} & \ding{55} & \ding{55} & \ding{55} \\
& Medications & \ding{51} & \ding{55} & \ding{51} & \ding{51}($\mathbb{T}$) & \ding{51}& \ding{55} \\
\midrule
Tasks & Labels & 15 & 770 & 23 & 5 & 9 & 33 \\
\midrule
Availability & Open source & \ding{51} & \ding{51} & \ding{51} & \ding{51}  & \ding{51}& \ding{51} \\
\bottomrule
\end{tabular}
\begin{tablenotes}
\item We use diverse symbology to express the contribution where \ding{51} = available, \ding{55} = unavailable, $\mathbb{E}$ = available in the form of embeddings, and $\mathbb{T}$ = available in the form of trends or at least two sampled values. $^{1}$HiRID only provides stay-level identifiers.
\end{tablenotes}
\end{table*}

\section{LLM Evaluation Prompts} 
\label{app:prompt}

\subsection{Prompt A: Initial Binary Prediction}

\begin{tcolorbox}[colback=gray!5,colframe=black!40,title=LLM Evaluation Prompt A]
You are an ICU expert and have to predict four binary clinical outcomes given ICU patient data:

1. Will the patient die during the ICU or hospital stay?
2. Will the patient require vasopressor administration within the next 24 hours?
3. Will the patient require invasive mechanical ventilation at any point within the next 24 hours?
4. Will the patient develop acute kidney injury within the next 24 hours, defined as either
   (a) a doubling of creatinine or
   (b) urine output below 500 ml?

Some features are binary e.g. gender, race, surgeries, and mechanical ventilations.
Other features are categorical like the GCS scores.
The rest is continuous.
for the features, we have either raw (in their respective unit of measurement) or differences (from one to another record). We also provide statistical features to observe trends e.g. min, max, first and last, and also the time in hours that has happened since that event until prediction time.

Here are the features and their units of measurements:
\texttt{\{Features\}}

This is the patient data:
\texttt{\{Patient data\}}

Please avoid excessive explanation, and just deliver binary predictions given the features.
\end{tcolorbox}

\subsection{Prompt B: Prediction Update After Model Probabilities}

\begin{tcolorbox}[colback=gray!5,colframe=black!40,title=LLM Evaluation Prompt B]
You are an ICU expert and have to predict four binary clinical outcomes given ICU patient data:

1. Will the patient die during the ICU or hospital stay?
2. Will the patient require vasopressor administration within the next 24 hours?
3. Will the patient require invasive mechanical ventilation at any point within the next 24 hours?
4. Will the patient develop acute kidney injury within the next 24 hours, defined as either
   (a) a doubling of creatinine or
   (b) urine output below 500 ml?

Some features are binary e.g. gender, race, surgeries, and mechanical ventilations.
Other features are categorical like the GCS scores.
The rest is continuous.
for the features, we have either raw (in their respective unit of measurement) or differences (from one to another record). We also provide statistical features to observe trends e.g. min, max, first and last, and also the time in hours that has happened since that event until prediction time.

Here are the features and their units of measurements:
\texttt{\{Features\}}

You have previously predicted this:
\texttt{\{Predictions\}}

For this patient data:
\texttt{\{Patient data\}}

You are now provided with predicted probabilities from a deep learning model.
\texttt{\{Probabilities\}}

Would you change any of your binary predictions after seeing these probabilities?
If so, provide an updated list of binary predictions.

Note: These probabilities do not imply a fixed 50\% decision threshold.
Avoid any explanation.
\end{tcolorbox}

\end{document}